\documentclass[12pt]{article}
\usepackage{graphicx}
\usepackage{subcaption}
\usepackage{mwe}
\usepackage{dcolumn}
\usepackage{bm}
\usepackage{epsfig}
\usepackage{color}
\usepackage{longtable}
\usepackage{setspace}
\usepackage{times}
\usepackage[a4paper,left=3cm,right=2cm,top=2cm,bottom=2cm,]{geometry}
\usepackage{authblk}
\usepackage{cite}

\def\ketm#1{  \left\vert  #1   \right\rangle   }

\def\sprm#1#2{  \left\langle #1 \left\vert \right. #2 \right\rangle   }

\def\mem#1#2#3{  \left\langle #1 \left\vert  #2 \right\vert #3 \right\rangle   }

\def\redmem#1#2#3{  \left\langle #1 \left\Vert
                  #2 \right\Vert #3 \right\rangle   }
\title{\vspace{-15mm}\selectfont\textbf{
Relativistic effects in the non-resonant two-photon $K$-shell ionization of neutral atoms
}} 

\author[1,2]{J.~Hofbrucker \thanks{Corresponding author: {j.hofbrucker.11@aberdeen.ac.uk}
}}
\author[2,3]{A.~V.~Volotka}
\author[1,2]{S.~Fritzsche}

\affil[1]{Theoretisch-Physikalisches Institut, Friedrich-Schiller-Universit\"at Jena, Max-Wien-Platz 1, D-07743
Jena, Germany}
\affil[2]{Helmholtz-Institut Jena, Fr\"o{}belstieg 3, D-07743 Jena, Germany}
\affil[3]{Department of Physics, St.~Petersburg State University, Oulianovskaya 1, 198504 St.~Petersburg, Russia}

\begin{document}
\maketitle
\doublespacing
\begin{abstract}

Relativistic effects in the non-resonant two-photon $K$-shell ionization of neutral atoms are studied theoretically within the framework of second-order perturbation theory. The non-relativistic results are compared with the relativistic calculations in the dipole and no-pair approximations as well as with the complete relativistic approach. The calculations are performed in both velocity and length gauges. Our results show a significant decrease of the total cross section for heavy atoms as compared to the non-relativistic treatment, which is mainly due to the relativistic wavefunction contraction. The effects of higher multipoles and negative continuum energy states counteract the relativistic contraction contribution, but are generally much weaker. While the effects beyond the dipole approximation are equally important in both gauges, the inclusion of negative continuum energy states visibly contributes to the total cross section only in the velocity gauge.

\end{abstract}

\section{Introduction}

The understanding of the limitations of non-relativistic theory applied to light-matter interaction has been of interest for many years. However, although theoretical studies were carried out, the experimental possibilities to verify the theoretical predictions were limited by the photon sources to low energies. This restriction has been overcome by the development of free electron lasers (FEL), which enable the production of intense photon beams with ultraviolet and x-ray energies \cite{Pellegrini/RMP:2016}. With such high-energy photon sources, the ionization of inner-shell electrons has become possible, and hence deep understanding of the theoretical approaches and their limitations is now required. 
Over the years, the one-photon one-electron ionization has become a well-studied process (see e.g. Refs. \cite{Oh/PRA:1976, Ron/PRA:1994, Demekhin/JPB:2014} and references therein). Furthermore, an extensive study of the sequential two-photon double ionization\cite{Fritzsche/JPB:2008, Fritzsche/JPB:2009, Kurka/JPB:2009, Fritzsche/JPB:2011} as well as general multiphoton single ionization (e.g. \cite{Lambropoulos/PRA:1974, Glushkov/JPC:2014} and references therein) have also been carried out. However, not much attention has been paid to (non-resonant) two-photon single-ionization of any general neutral atom.
Two-photon ionization (TPI) of a single electron is one of the fundamental non-linear processes in the light-matter interaction, which offers different selection rules and the possibility of ionization of heavier atoms in comparison to the one-photon ionization. 

The first TPI experiments utilizing the FEL facilities were carried for the ionization of the $4d$ electron of neutral Xe atom \cite{Richardson/PRL:2010}, $1s$ electrons of Ne$^{8+}$ ion \cite{Doumy/PRL:2011} and He atom \cite{Ma/JPB:2013}. In all of these experiments, either an electron or an ion spectrometer was used to detect the TPI process. However, in the $K$-shell ionization of neutral atoms, these detection techniques may not be convenient due to the small cross sections of TPI. More promising method to study the TPI process is to detect the $K$-fluorescence photons, which serve as a direct signature of the $K$-shell vacancy. This experimental approach has been utilized in the measurements of the $K$-shell TPI of neutral Ge \cite{Tamasaku/NP:2014}, Zr \cite{Ghimire/PRA}, and Cu \cite{Szlachetko/SR:2016} atoms.

Theoretically, TPI was studied in detail already 50 years ago, when the first non-relativistic calculations of the TPI cross section of atomic hydrogen were carried out and presented together with the well-known $Z^{-6}$ scaling law ($Z$ is the nuclear charge number) for any other hydrogenlike ion \cite{Zernik/PR:1964}. However, it was shown later \cite{Koval/JPB:2003, Koval/JPB:2004, Koval/Dissertation}, that a rather essential deviation from the scaling law occurs due to the relativistic effects. Recently, the retardation effects in the above-threshold TPI of low-$Z$ hydrogenlike ions were also investigated in Refs. \cite{Florescu/PRA:2011, Florescu/PRA:2012}. Although a significant difference from the scaling law was found for hydrogenlike ions, no systematic study has been performed for the TPI of neutral atoms until now. In our recent work \cite{Hofbrucker/PRA:2016}, we have shown that the screening potential created by the electrons of the neutral atom leads to a minimum in the non-resonant TPI cross section near the ionization threshold, which is absent for hydrogenlike ions. Moreover, we have investigated the deviation from the scaling law due to both the screening as well as relativistic effects. It is the purpose of this work to go a step further and explicitly separate the individual contributions of relativistic effects and enumerate their corresponding strengths.

In Sec. \ref{Sec_theory}, we present a brief description of the applied theoretical formalism. Section \ref{Sec_results} discusses the importance of relativistic effects; relativistic wavefunction contraction, inclusion of higher multipoles, and summation over the negative continuum energy states. Finally, a summary is given in Sec. \ref{Sec_summary}. Relativistic units ($\hbar=c=m=1$) are used throughout the paper, unless stated otherwise.

\section{Theory}\label{Sec_theory}

We shall not provide a detailed derivation of the total cross section which is presented already in Ref. \cite{Hofbrucker/PRA:2016}, but restrict  only to the formulae needed for general understanding and further discussion. Let us consider the non-resonant two-photon one-electron ionization process, where the two photons are assumed to be identical, i.e. with equal wave and polarization vectors $\bm{k}$ and $\bm{\hat{\varepsilon}}_{\lambda}$, respectively. This corresponds to the most common experimental setup, where the two photons originate from the same source. This process can be expressed as
\begin{eqnarray}
\ketm{\alpha_i J_i M_i}+2\gamma(\bm{k},\bm{\hat{\varepsilon}}_{\lambda}) \rightarrow \ketm{\alpha_f J_f M_f}+\ketm{\bm{p}_e m_e},
\end{eqnarray}
where the atom is initially in the many-electron state $\ketm{\alpha_i J_i M_i}$ with the total angular momentum $J_i$, its projection $M_i$, and where $\alpha_i$ denotes all further quantum numbers necessary for unique characterization of the state. After simultaneous absorption of two identical photons $\gamma(\bm{k},\bm{\hat{\varepsilon}}_{\lambda})$ with energies $\omega=\bm{k}/|\bm{k}|$, the system consists of a singly charged ion $\ketm{\alpha_f J_f M_f}$ with quantum numbers $\alpha_f, J_f, M_f$ characterizing the final state and a free electron in a state $\ketm{\bm{p}_e m_e}$ with well-defined asymptotic momentum $\bm{p}_e$ and a spin projection $m_e$. Using the density matrix theory, we can describe the final state of our system, in terms of the density matrices of the initial system $\mem{\alpha_i J_i M_i, \bm{k}\lambda_1 \bm{k}\lambda_2}{\hat{\rho}}{\alpha_i J_i M_i', \bm{k}\lambda_1' \bm{k}\lambda_2'}$ and the transition amplitude $M^{\lambda_1 \lambda_2}_{J_i M_i J_f M_f m_e}$, which describes the electron-photon interaction. As the atom and the incident radiation are initially independent, the initial-state density matrix can be written as $\mem{\alpha_i J_i M_i, \bm{k}\lambda_1 \bm{k}\lambda_2}{\hat{\rho}}{\alpha_i J_i M_i',\bm{k}\lambda_1'\bm{k}\lambda_2'}=\mem{\bm{k}\lambda_1}		{\hat{\rho}_{\gamma}}{\bm{k}\lambda_1'}\mem{\bm{k}\lambda_2}{\hat{\rho}_{\gamma}}{\bm{k}\lambda_2'} \mem{\alpha_i J_i M_i}{\hat{\rho}_i}{\alpha_i J_i M_i'}$. Under the assumption of an initially unpolarized neutral atom, the corresponding density matrix simplifies to $\mem{\alpha_i J_i M_i}{\hat{\rho}_i}{\alpha_i J_i M_i'}=1/[J_i] \delta_{M_i M_i'}$. The trace of the final density matrix gives us the TPI cross section
\begin{eqnarray}
\label{TotalCrossSection}
\sigma(\omega)&=& \frac{32 \pi^5 \alpha^2}{\omega^2}
	\sum_{ \lambda_1 \lambda_2 \lambda_1' \lambda_2'}
	\mem{\bm{k}\lambda_1}
		{\hat{\rho}_{\gamma}}
		{\bm{k}\lambda_1'}
	\mem{\bm{k}\lambda_2}
		{\hat{\rho}_{\gamma}}
		{\bm{k}\lambda_2'}\\\nonumber &\times&
	\frac{1}{[J_i]}\sum_{J_f M_f M_i m_e} \int d\Omega_{\bm{\hat{p}}_e}
	M^{\lambda_1 \lambda_2}_{J_i M_i J_f M_f m_e}
	M^{\lambda_1' \lambda_2' *}_{J_i M_i J_f M_f m_e}\\\nonumber
&=&\frac{32 \pi^5 \alpha^2}{\omega^2}
	\sum_{ \lambda_1 \lambda_2 \lambda_1' \lambda_2'}
	\mem{\bm{k}\lambda_1}
		{\hat{\rho}_{\gamma}}
		{\bm{k}\lambda_1'}
	\mem{\bm{k}\lambda_2}
		{\hat{\rho}_{\gamma}}
		{\bm{k}\lambda_2'}
	\sum_{m_a j l m_j}
	T^{\lambda_1 \lambda_2}_{m_a j l m_j}
	T^{\lambda_1' \lambda_2'*}_{m_a j l m_j},
\end{eqnarray}
where we presumed that the photoelectrons are detected in 4$\pi$ solid angle but their polarization is not observed, therefore we integrated over the directions of the emitted electron $\Omega_{\bm{\hat{p}}_e}$ and summed over the spin projection $m_e$. As the final state of the ion is not observed, summations over $J_f$ and $M_f$ have been carried out as well. The photon helicity density matrices $\mem{\bm{k}\lambda}{\hat{\rho}_{\gamma}}{\bm{k}\lambda'}$ allow us to conveniently parametrize the polarization of the photons by means of the linear ($P_1, P_2$) and circular ($P_3$) Stokes parameters.
The second equality in Eq. (\ref{TotalCrossSection}) expresses the total cross section in terms of so called angle-reduced transition amplitude $T^{\lambda_1 \lambda_2}_{m_a j l m_j}$ which comes from further simplifications of the general many-electron transition amplitude $M^{\lambda_1 \lambda_2}_{J_i M_i J_f M_f m_e}$ within independent-particle approximation. The general amplitude can be represented in second-order perturbation theory as

\begin{eqnarray}
\label{GeneralTransitionAmplitude}
M_{J_i M_i J_f M_f m_e}^{\lambda_1 \lambda_2}=\int\kern-1.2em\sum_{~\nu}\frac{
	\mem{\alpha_fJ_fM_f,\bm{p}_e m_e}
		{\hat{\mathcal{R}}(\bm{k},\bm{\hat{\varepsilon}}_{\lambda})}
		{\alpha_{\nu}J_{\nu}M_{\nu}}
	\mem{\alpha_{\nu}J_{\nu}M_{\nu}}
		{\hat{\mathcal{R}}(\bm{k},\bm{\hat{\varepsilon}}_{\lambda})}
		{\alpha_iJ_iM_i}		
		}
	{E_{i}+\omega-E_{\nu}}, 
\end{eqnarray}
where ${\hat{\mathcal{R}}(\bm{k},\bm{\hat{\varepsilon}}_{\lambda})}$ is the one-particle transition operator, $E_i$ and $E_\nu$ are the energies of the initial and intermediate many-electron states. 
We can simplify this expression by applying the independent-particle approximation and the particle-hole formalism. In this framework, we can describe the TPI process as follows. The simultaneous absorption of the two photons by the neutral atom ejects an electron from an atomic subshell $\ketm{n_a j_a l_a m_a}$ via virtual intermediate state $\ketm{n_n j_n l_n m_n}$ into a continuum state $\ketm{\bm{p}_e m_e}$, leaving a hole (or vacancy) behind. The $n, j, l$, and $m$ describe the one-electron principle, total angular momentum, orbital angular momentum, and projection of the total angular momentum quantum numbers, respectively. According to the particle-hole formalism, the final ionized state of the atom can be described by applying the hole creation operator to the initial state and coupling the corresponding angular momenta. By carrying out this simplification, the many-electron transition amplitude simplifies to an amplitude depending only on the one-electron wavefunctions of the active electron. Furthermore, we expand the continuum electron wavefunction into partial waves and carry out the multipole expansion of the photon wavefunction. Then, using the Wigner-Eckart theorem, we obtain an expression depending on the reduced matrix elements, which describe the electron-photon interaction independently of the magnetic quantum numbers $m_a$, $m_n$, and $m_e$. Finally, the second expression of the total cross section (\ref{TotalCrossSection}) was obtained by carrying out the integration over the 4$\pi$ solid angle $\Omega_{\bm{\hat{p}}_e}$ and the summation over electron spin projection $m_e$. By performing all the above steps, the angle-reduced transition amplitude from Eq. (\ref{TotalCrossSection}) can be written in the independent-particle approximation as follows

\begin{eqnarray}\label{ReducedMatrixElement}
T^{\lambda_1 \lambda_2}_{m_a j l m_j}&=& 
	\sum_{p_1 J_1} \sum_{p_2 J_2} \sum_{n_n j_n l_n m_n}
	i^{J_1-p_1+J_2-p_2} 	\sqrt{\frac{[J_1, J_2]}{[j_n, j_a]}} 
	(-\lambda_1)^{p_1} (-\lambda_2)^{p_2}\\\nonumber
	&\times&
	(-1)^{j-m_j} 
	\sprm{j,m_j,J_1,-\lambda_1}{j_n,m_n}
	\sprm{j_n, m_n, J_2,-\lambda_2}{j_a, m_a}\\\nonumber
	 &\times&\frac{
		\redmem{\varepsilon_e j l}
				{\bm{\alpha} \cdot \bm{ a}^{(p_1)}_{J_1}}
				{n_n j_n l_n}
		\redmem{n_n j_n l_n}
				{\bm{\alpha} \cdot \bm{ a}^{(p_2)}_{J_2}}
				{n_a j_a l_a}
		}{E_{n_a j_a l_a}+\omega-E_{n_n j_n l_n}},
\end{eqnarray}
where $[J]=2J+1$, $\sprm{. . . .}{. .}$ represents a Clebsch-Gordan coefficient, $\varepsilon_e$ is the electron energy, $j$, $l$, and $m_j$ are the angular momentum quantum numbers of the continuum electron, $J$ and $M$ are the quantum numbers of the photon multipoles and the index $p$ describes the electric ($p=1$) and magnetic ($p=0$) components of the photon wavefunction. Note, that the angle-reduced transition amplitude $T^{\lambda_1 \lambda_2}_{m_a j l m_j}$ is completely independent of the many-electron state quantum numbers. A more general expression of the transition amplitude for the two-photon ionization can be found in Ref. \cite{Hofbrucker/PRA:2016}.

The results presented in the following section are obtained by solving the Dirac equation with Core-Hartree screening potential. In Ref. \cite{Hofbrucker/PRA:2016}, we have shown that there is no significant dependence of the total cross section nor the relativistic effects on the choice of screening potential.
However, we also showed, that the account for the other electrons can lead to a strong decrease of the dominant ionization channel. The electron correlations in radiative transitions have also been studied e.g. in \cite{Glushkov/PLA:1992, Fink/PRA:1990, Novikov/JPB:2001}.
 To sum over the infinite number of intermediate states, finite basis set \cite{Sapirstein/JPB:1996} constructed from $B$-splines by applying the dual-kinetic-balance approach \cite{Shabaev/PRL:2004} is employed. This approach has been previously successfully applied, for example, in the calculations of two-photon decay rates \cite{Surzhykov/PRA:2010 , Volotka/PRA:2011} and Rayleigh scattering \cite{Volotka/PRA:2016} in heliumlike ions . The continuum-state wavefunctions are obtained numerically by solving the Dirac equation with the help of the RADIAL package \cite{Salvat/CPC:1995}.

\section{Results and discussion}\label{Sec_results}

\begin{figure}[t!]
        \centering
        \begin{subfigure}[b]{0.48\textwidth}
            \centering
            \includegraphics[width=\textwidth]{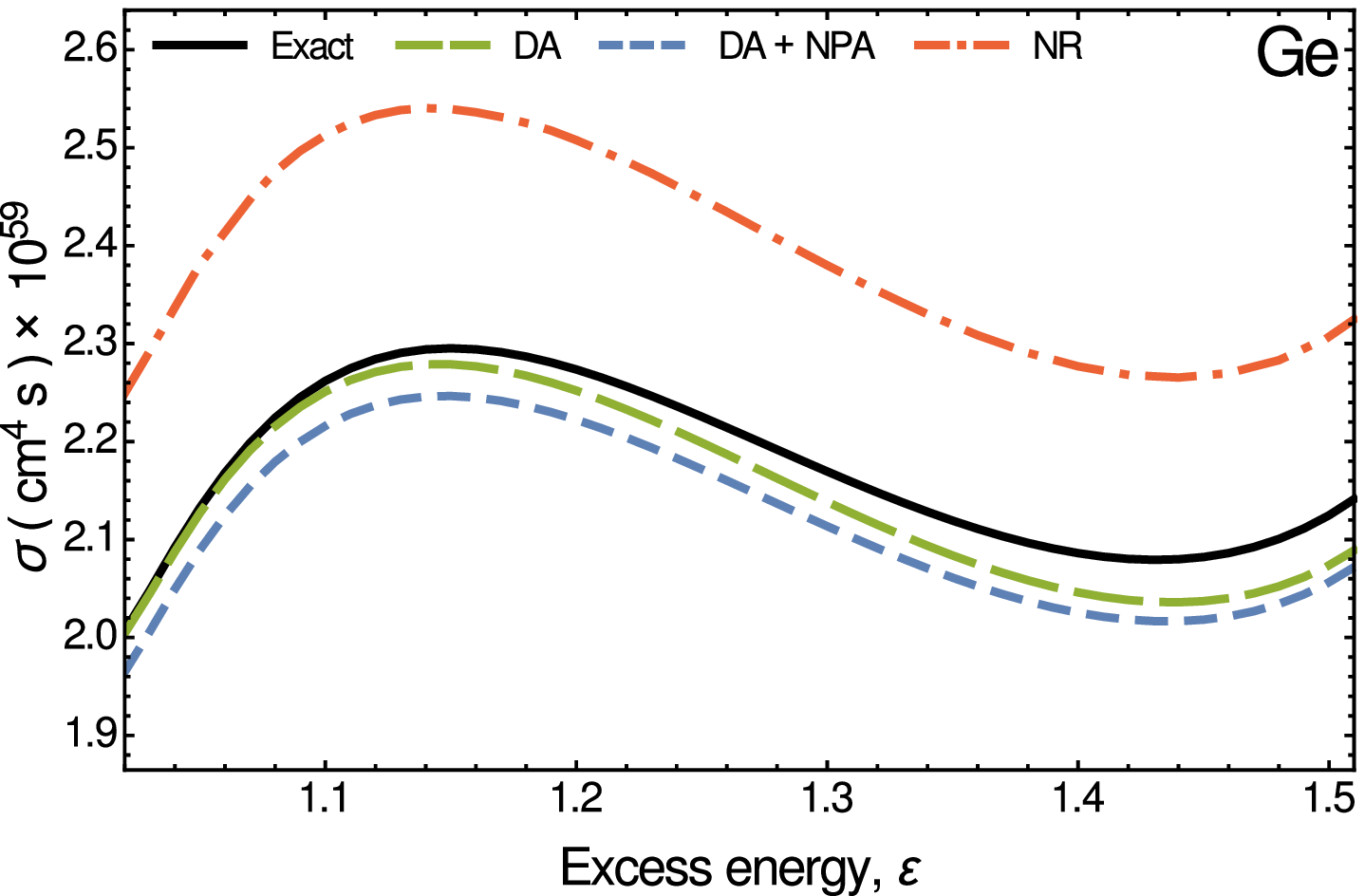}
        \end{subfigure}
        \hfill
        \begin{subfigure}[b]{0.48\textwidth}  
            \centering 
            \includegraphics[width=\textwidth]{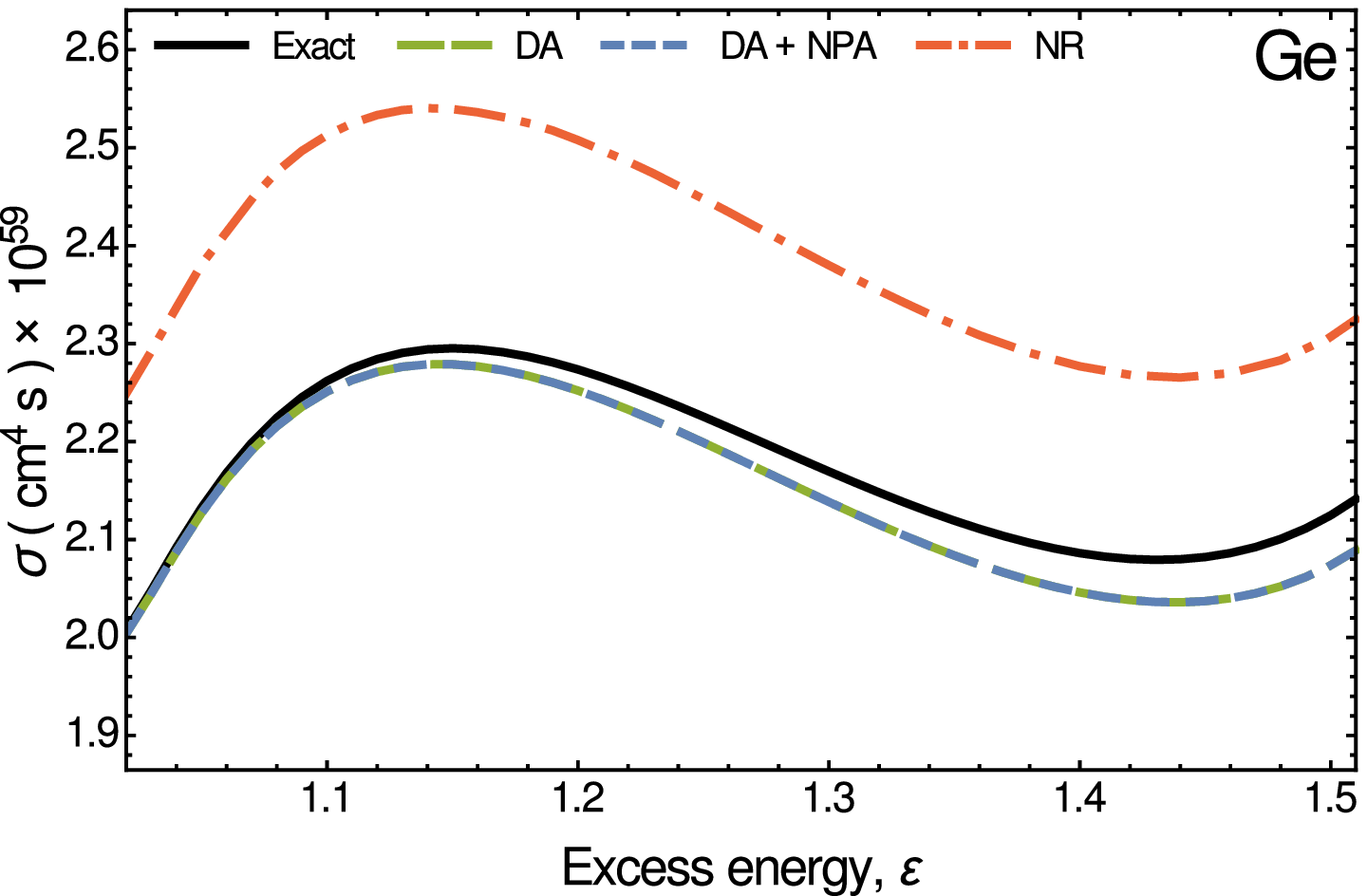}
        \end{subfigure}
        \vskip\baselineskip
        \begin{subfigure}[b]{0.48\textwidth}   
            \centering 
            \includegraphics[width=\textwidth]{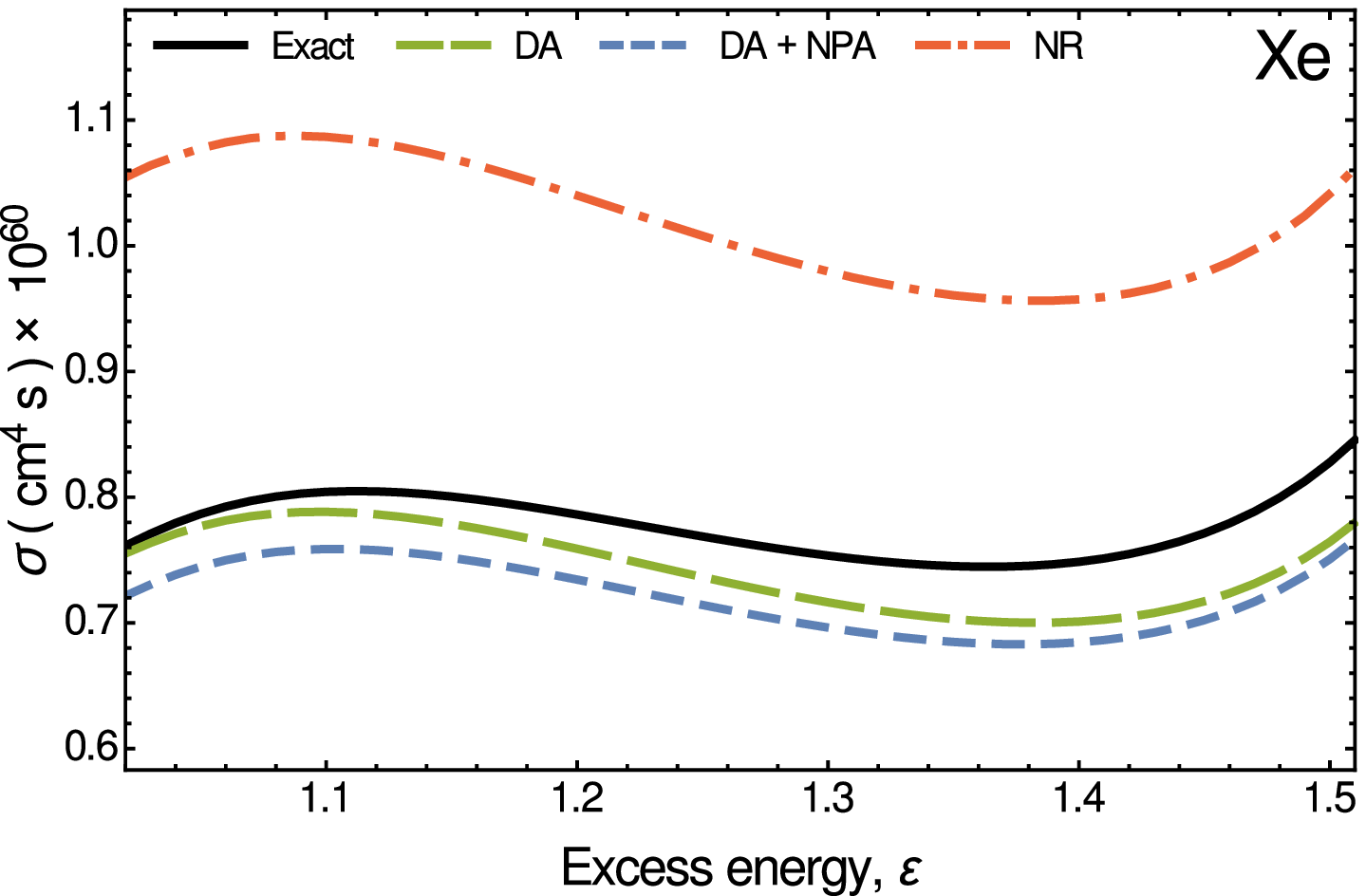}
        \end{subfigure}
        \quad
        \begin{subfigure}[b]{0.48\textwidth}   
            \centering 
            \includegraphics[width=\textwidth]{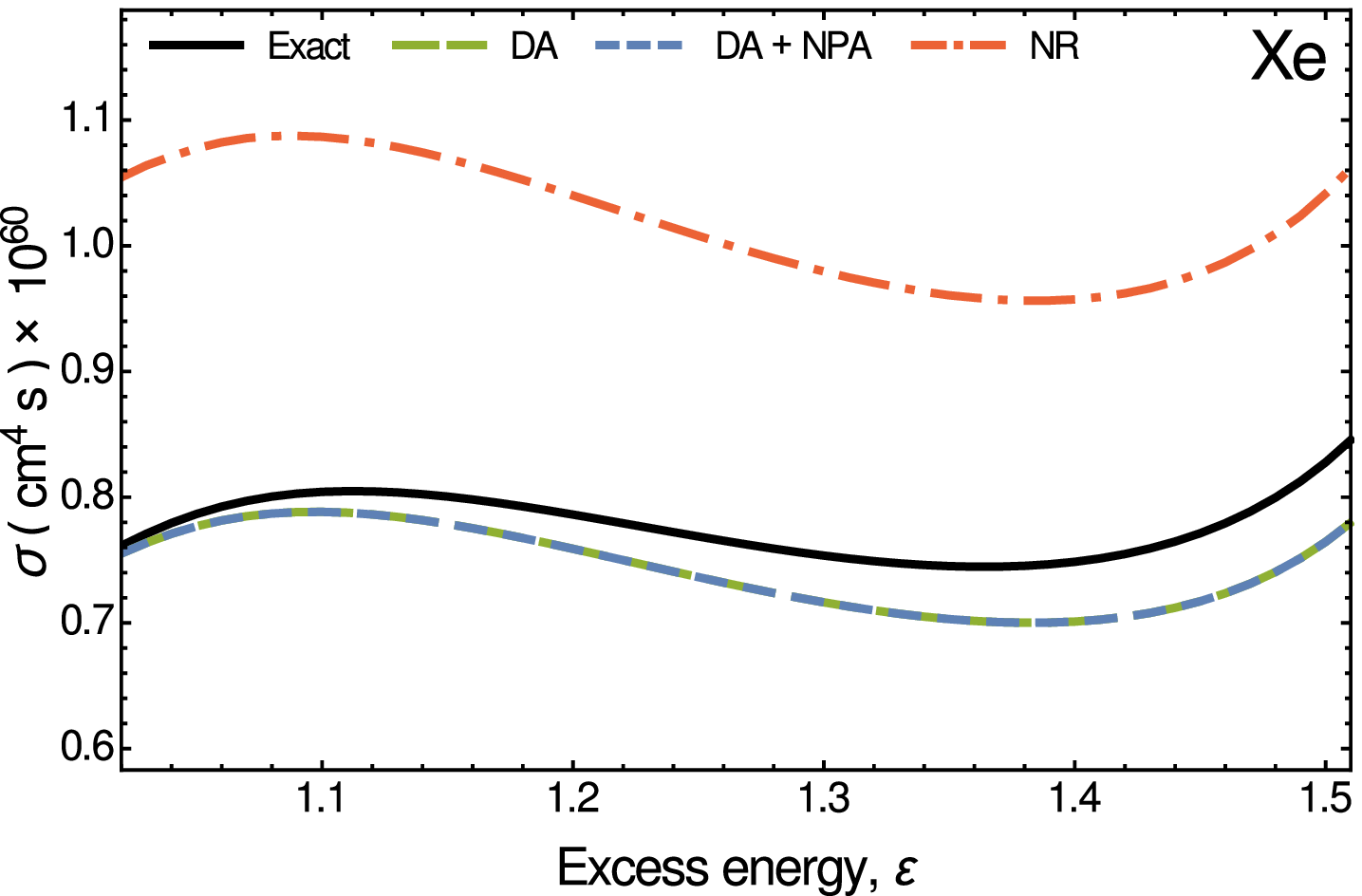}
        \end{subfigure}
                \vskip\baselineskip
        \begin{subfigure}[b]{0.48\textwidth}   
            \centering 
            \includegraphics[width=\textwidth]{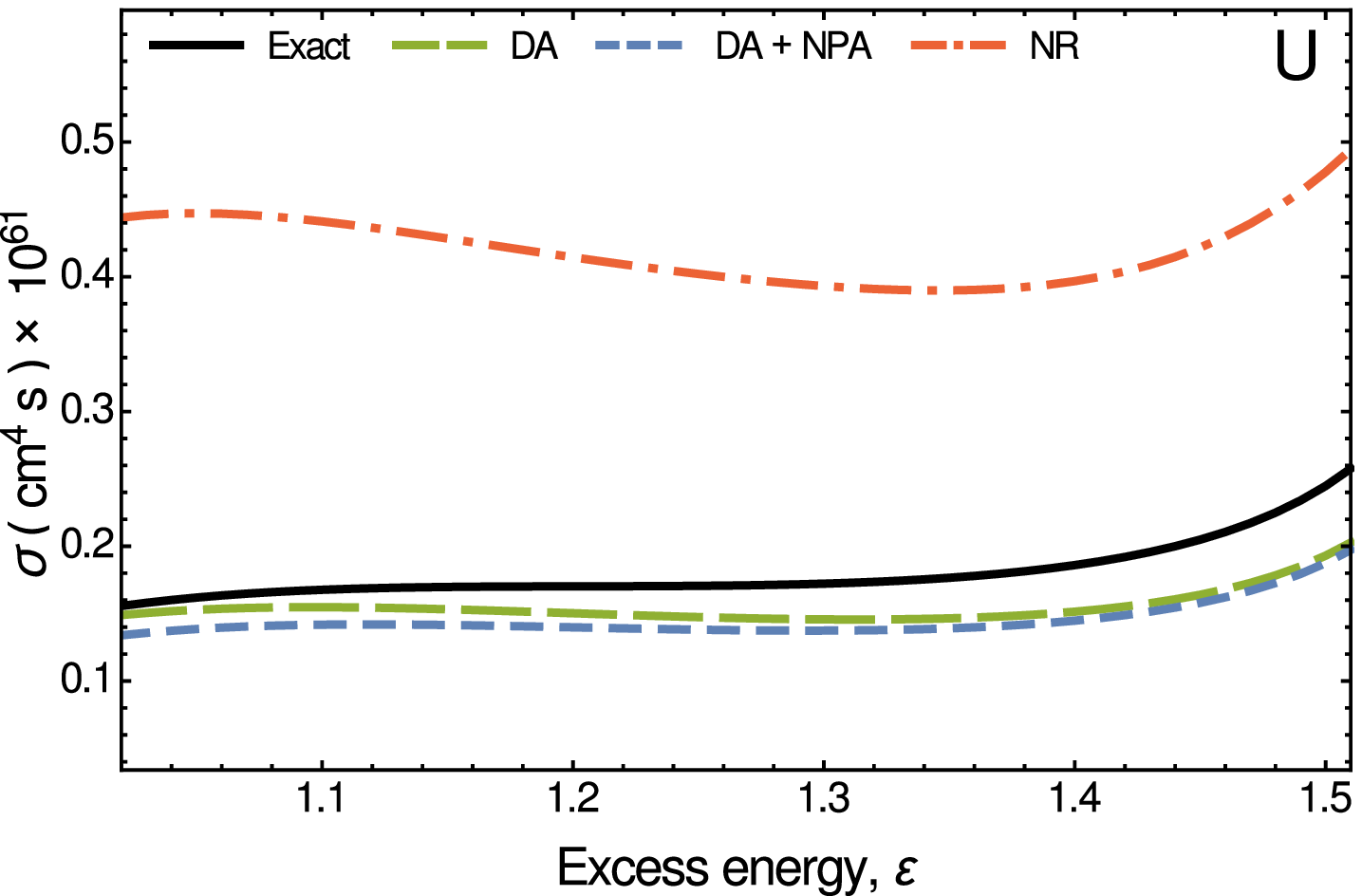} 
            \caption[]%
            {{Velocity gauge}}  
        \end{subfigure}
        \quad
        \begin{subfigure}[b]{0.48\textwidth}   
            \centering 
            \includegraphics[width=\textwidth]{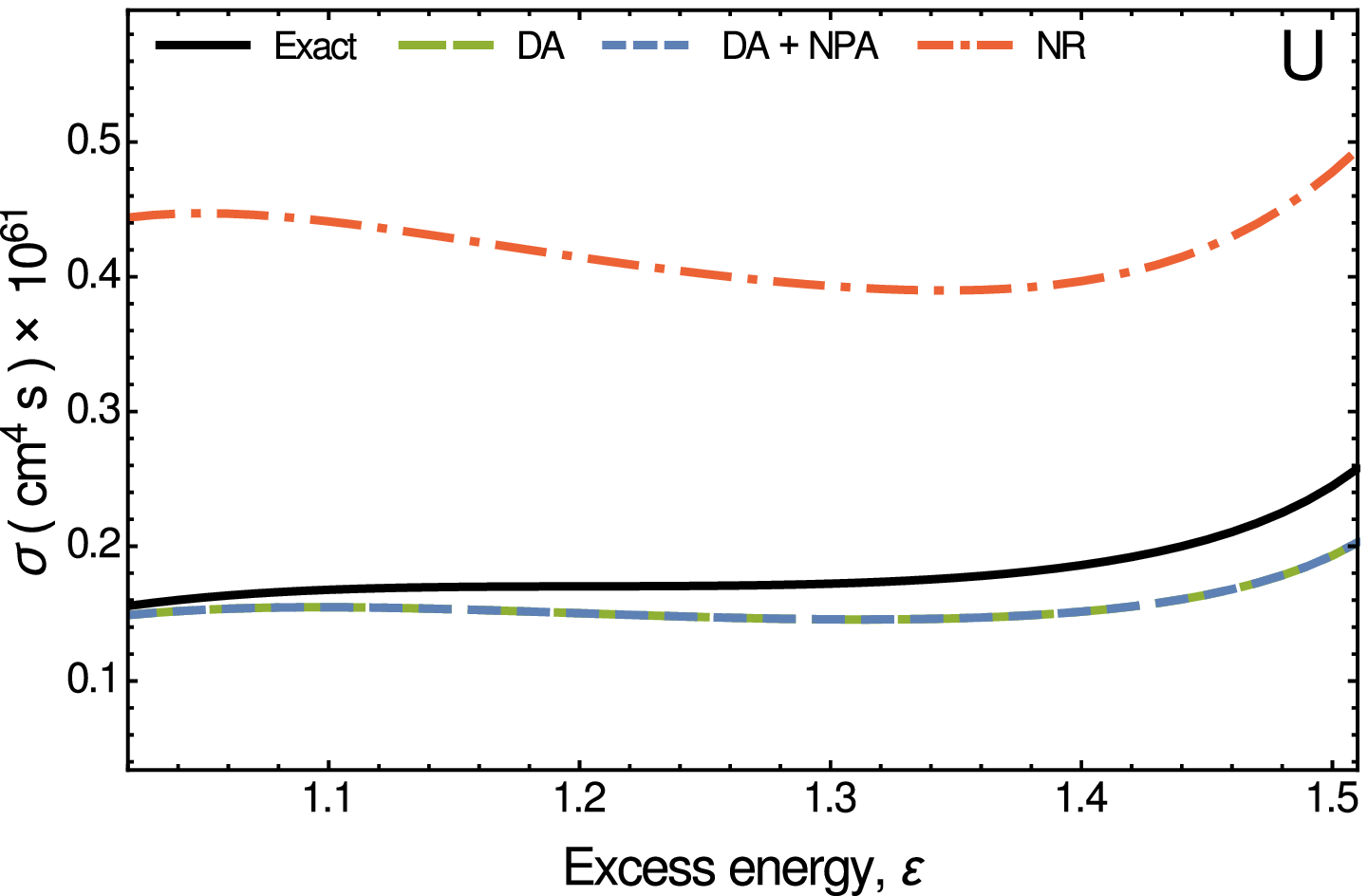}
            \caption[]%
            {{Length gauge}}    
        \end{subfigure}
        \caption[]
        {Total non-resonant $K$-shell two-photon ionzation cross section $\sigma$ as a function of excess energy within different approximations; exact relativistic $\sigma^{\mathrm{Exact}}$ (solid black), dipole $\sigma^{\mathrm{DA}}$ (long-dashed green), dipole + no-pair $\sigma^{\mathrm{DA + NPA}}$ (short-dashed blue), and non-relativistic $\sigma^{\mathrm{NR}}$ (dot-dashed red). The calculations are carried out in (a) velocity (left column) and (b) length (right column) gauges for ionization of neutral germanium, xenon, and uranium atoms.}\label{Fig_cross_sections}
    \end{figure}
    
Calculations of the TPI cross section can be further performed within different approximations in order to investigate the importance of various effects. Each of the approximations can be understood as a certain simplification of Eq. (\ref{ReducedMatrixElement}). First, in order to study the effects of higher-order multipoles, we restrict the infinite summations over the multipoles $pJ$ to $p=1$ (electric) and $J=1$ (dipole) terms only. This approximation is known as the dipole approximation (DA), and we denote the corresponding cross section as $\sigma^{\mathrm{DA}}$. Moreover, the summation in Eq. (\ref{ReducedMatrixElement}) over the virtual intermediate states $\ketm{n_n j_n l_n m_n}$ runs over the complete (positive and negative) energy spectrum. The presence of negative-energy states in the sum corresponds to the process with creation of a positron in the intermediate state. Thus, in order to enumerate the contribution from this process, we, in addition to the DA, restrict summation over the intermediate states to the positive energy states only. We refer to this calculation as dipole and no-pair approximations (DA + NPA), and denote the corresponding cross section as $\sigma^{\mathrm{DA + NPA}}$. Finally, we consider also the non-relativistic limit (NR) of Eq. (\ref{ReducedMatrixElement}). For this, we employ the wavefunctions which are the solutions of the Sch\"odinger equation and replace interaction operators $\bm{\alpha}\cdot \bm{a}^{(p)}_{J}$ by its non-relativistic limit $\omega r / \sqrt{6\pi}$ and also set $p_1=p_2=1$ and $J_1=J_2=1$. The corresponding cross section is denoted as $\sigma^{\mathrm{NR}}$.
If, however, no approximation is made, i.e., the Dirac equation is used to obtain the electron wavefunctions, summation over the intermediate states runs over both positive and negative energy states, and all multipoles are taken into account, we refer to such calculations as "Exact", and we write the cross sections as $\sigma^{\mathrm{Exact}}$. Actually, the multipole summation is restricted to $J_{\mathrm{max}}=5$, which is sufficient to obtain convergence of the corresponding total cross section at less than $0.001\%$ level.

\begin{figure}[t]
		\centering
        \begin{subfigure}[b]{0.475\textwidth}   
            \centering 
            \includegraphics[width=\textwidth]{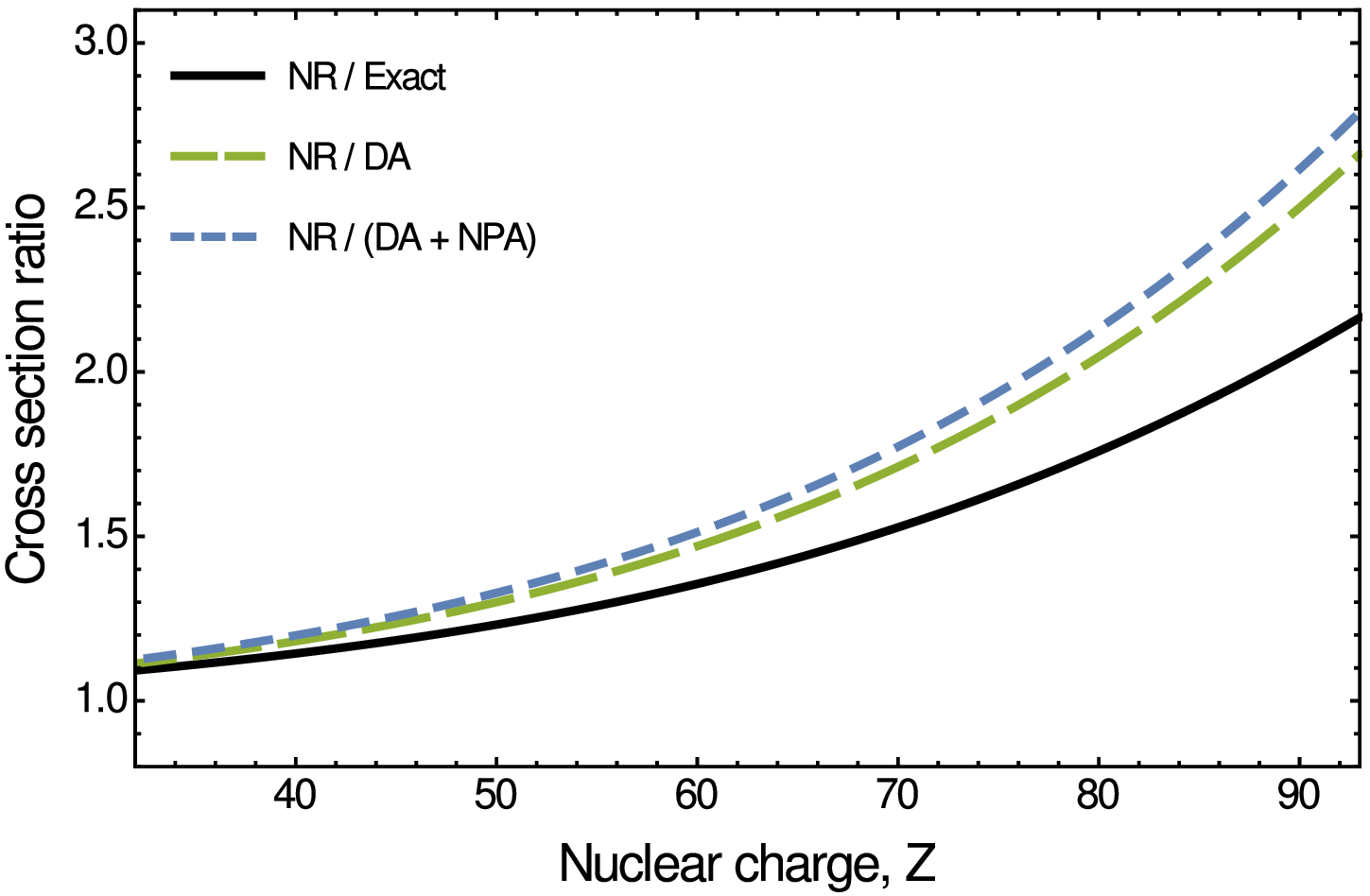}
            \caption[]%
            {{Velocity gauge}}
        \end{subfigure}
        \quad
        \begin{subfigure}[b]{0.475\textwidth}   
            \centering 
            \includegraphics[width=\textwidth]{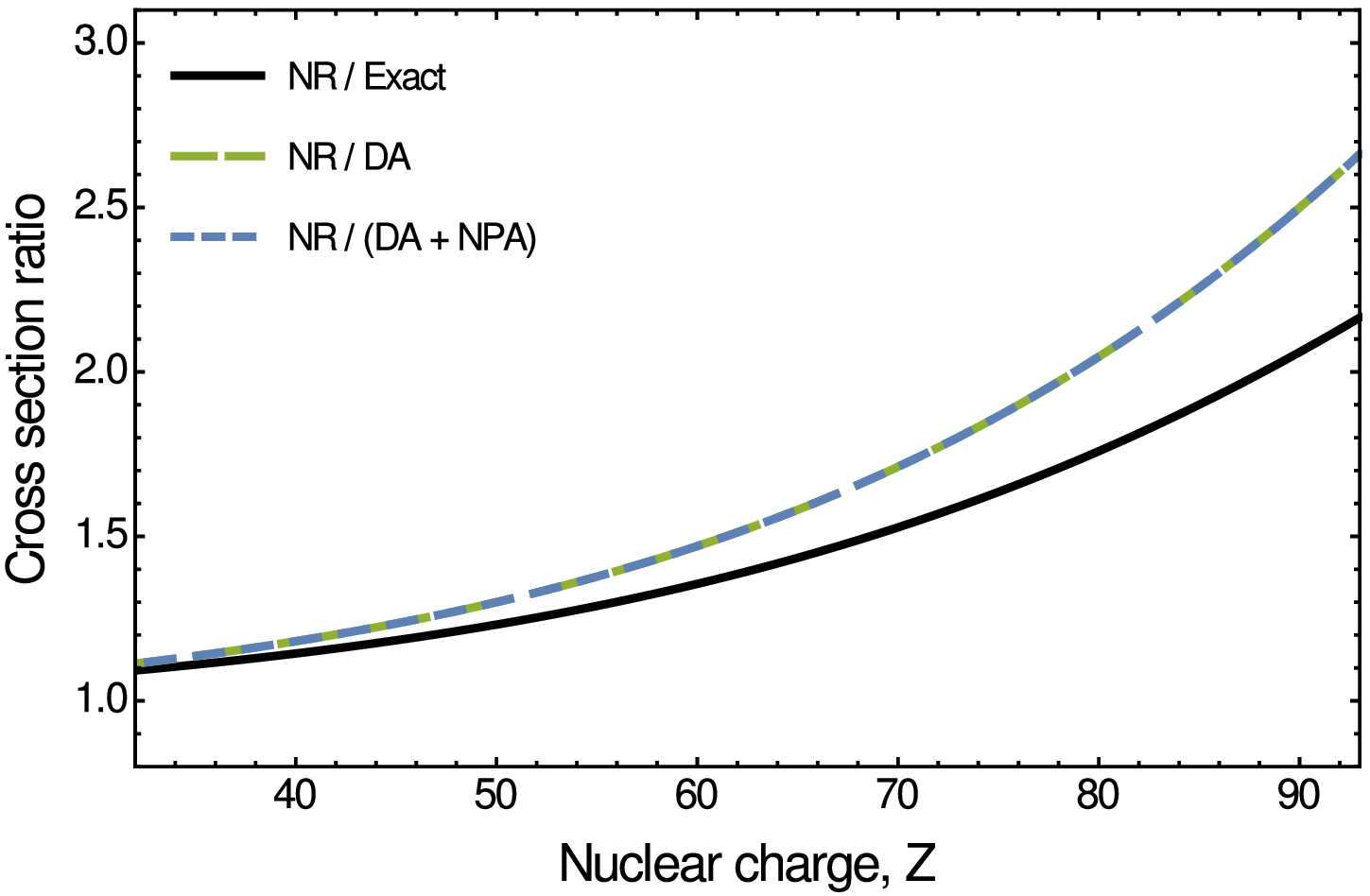}
            \caption[]%
            {{Length gauge}}
        \end{subfigure}
        \caption[]
        {The cross section ratio as a function of nuclear charge; $\sigma^{\mathrm{NR}}/\sigma^{\mathrm{Exact}}$ (solid black), $\sigma^{\mathrm{NR}}/\sigma^{\mathrm{DA}}$ (long-dashed green), and $\sigma^{\mathrm{NR}}/\sigma^{\mathrm{DA+NPA}}$ (short-dashed blue) in (a) velocity and (b) length gauges. The results correspond to $\varepsilon=1.40$ excess energy.} 
        \label{Fig_ratio}
    \end{figure}

Figure \ref{Fig_cross_sections} presents the total non-resonant $K$-shell TPI cross section as function of excess energy for the ionization of neutral Ge, Xe, and U atoms by linearly polarized light. Excess energy is the combined two-photon energy in units of the ionization threshold energy $E_{\mathrm{bind}}$, i.e., $\varepsilon=2\omega/E_{\mathrm{bind}}$. The minima in the total cross sections (see Fig. \ref{Fig_cross_sections}) in near-threshold energies occur as consequences of screening effects. For more details, we refer the reader to our previous work \cite{Hofbrucker/PRA:2016}. Here, we compare the total cross section values within various approximations and see that the major difference is present between $\sigma^{\mathrm{NR}}$ and all other approximations. The reason for this is that the Dirac wavefunctions have been used in all calculations, except the NR one, and solving the Dirac equation results in a contraction of the electron wavefunction. As a consequence of this contraction, the total TPI cross section is significantly lower in the relativistic description. We would expect that the decrease of the exact calculation (in comparison to the NR limit) should be "stronger" with increasing nuclear charge and photon energy. However, while it is true that the cross section drop increases with nuclear charge, it slowly decreases with energy. This is due to the higher multipole (beyond DA) effects, which open further channels for the ionization. As it is clear from Fig. \ref{Fig_cross_sections}, the cross section values in DA coincide with the exact calculation for near threshold energies, however, the "strength" of multipole effects increases with energy and counteracts the cross section decrease due to wavefunction contraction. Thus, in the exact calculation, the "strength" of the overall relativistic effects slowly decreases with energy. We can see that this is the case both in velocity as well as length gauges. 
The gauge-independence does not hold any longer for the NPA. Our results show that DA+NPA calculations result in a decrease of the total cross section values in the velocity gauge, while in the length gauge they result only in negligible effect (less than 0.05\%). Thus, the negative continuum energy effects are only essential in the velocity gauge, where they lead to an increase of the cross section by up to 10$\%$ as compared to the length gauge. The strong gauge-dependence of negative continuum energy effects has been previously also reported for the case of two-photon bound-bound transitions in hydrogenlike ions \cite{Labzowsky/JPB:2005 , Surzhykov/PRA:2009}.

In order to enumerate the importance of relativistic effects as a function of nuclear charge, we compare the calculations discussed above to the non-relativistic approximation, by introducing the ratio $\sigma^{\mathrm{NR}}/\sigma$, where $\sigma$ represents a relativistic evaluation either in DA, DA+NPA, or the exact calculation. Figure \ref{Fig_ratio} presents such ratios as a function of nuclear charge in both velocity and length gauges. This figure displays explicitly that all relativistic effects increase with nuclear charge, and it also shows that the negative continuum effects result in no significant effect in the length gauge across all nuclear charges. In general, the figure demonstrates the importance of relativistic effects, as a result, we stress that in order to obtain an agreement with future experiment, the relativistic effects need to be taken account for heavier atoms, for which the cross section drops by up to a factor 3 in comparison to the nonrelativistic prediction. Our results are in a good agreement with available experimental data in the designated energy range \cite{Tamasaku/NP:2014, Ghimire/PRA}, see \cite{Hofbrucker/PRA:2016} for more elaborative comparison.

\section{Summary}\label{Sec_summary}

In summary, relativistic calculations of the total non-resonant $K$-shell two-photon ionization have been performed. These results have been compared to calculations in three different approximations; dipole, no-pair, and non-relativistic. It has been shown that the importance of inclusion of the relativistic effects grows with increasing nuclear charge and the main contribution to the effects arises from the relativistic wavefunction contraction. The contributions from higher multipoles and negative continuum energy states increase the cross section, however, they are generally much smaller than the relativistic contraction.

\section*{Acknowledgements}
This work has been supported by the BMBF (Grant No. 05K13VHA).



\begin{thebibliography}{99} 

\bibitem{Pellegrini/RMP:2016} C. Pellegrini, A. Marinelli, and S. Reiche, Rev. Mod. Phys. 88 (2016) 015006.

\bibitem{Oh/PRA:1976} S. D. Oh, J. McEnnen, and R. H. Pratt, Phys. Rev. A 14 (1976) 1428.

\bibitem{Ron/PRA:1994} Akiva Ron, \textit{et al.}, Phys. Rev. A 50 (1994) 1312.

\bibitem{Demekhin/JPB:2014} Ph. V. Demekhin, J. Phys. B 47 (2014) 025602.

\bibitem{Lambropoulos/PRA:1974} P. Lambropooulos, Phys. Rev. A 9 (1974) 1992.

\bibitem{Glushkov/JPC:2014} A. V. Glushkov, J. Phys.: Conf. Ser. 548 (2014) 012020.

\bibitem{Fritzsche/JPB:2008} S. Fritzsche, A. N. Grum-Grzhimailo, E. V. Gryzlova, and N. M. Kabachnik, J. Phys. B 41 (2008) 165601.

\bibitem{Fritzsche/JPB:2009} S. Fritzsche, A. N. Grum-Grzhimailo, E. V. Gryzlova, and N. M. Kabachnik, J. Phys. B 42 (2009) 145602.

\bibitem{Kurka/JPB:2009} M. Kurka, \textit{et al.}, J. Phys. B 42 (2009) 141002. 

\bibitem{Fritzsche/JPB:2011} S. Fritzsche, A. N. Grum-Grzhimailo, E. V. Gryzlova, and N. M. Kabachnik, J. Phys. B 44 (2011) 175602.

\bibitem{Richardson/PRL:2010} V. Richardson, \textit{et al.}, Phys. Rev. Lett. 105 (2010) 013001.

\bibitem{Doumy/PRL:2011} G. Doumy, \textit{et al.}, Phys. Rev. Lett. 106 (2011) 083002.

\bibitem{Ma/JPB:2013} R. Ma, \textit{et al.}, J. Phys. B 46 (2013) 164018.

\bibitem{Tamasaku/NP:2014} K. Tamasaku, \textit{et al.}, Nat. Photon. 8 (2014) 313.

\bibitem{Ghimire/PRA} S. Ghimire, \textit{et al.}, Phys. Rev. A. 94 (2016) 043418.

\bibitem{Szlachetko/SR:2016} J. Szlachetko, \textit{et al.}, Sci. Rep. 6 (2016) 33292.

\bibitem{Zernik/PR:1964} W. Zernik, Phys. Rev. 135 (1964) A51.

\bibitem{Koval/JPB:2003} P. Koval, S. Fritzsche, and A. Surzhykov, J. Phys. B 36 (2003) 873.

\bibitem{Koval/JPB:2004} P. Koval, S. Fritzsche, and A. Surzhykov, J. Phys. B 37 (2004) 375388.

\bibitem{Koval/Dissertation} P. Koval, Two-photon ionization of atomic inner-shells, University of Kassel, (2004).

\bibitem{Florescu/PRA:2011} V. Florescu, O. Budriga, and H. Bachau, Phys. Rev. A 84 (2011) 033425.

\bibitem{Florescu/PRA:2012} V. Florescu, O. Budriga, and H. Bachau, Phys. Rev. A 86 (2012) 033413.

\bibitem{Hofbrucker/PRA:2016} J. Hofbrucker, A. V. Volotka, and S. Fritzsche, Phys. Rev. A 94 (2016) 063412.

\bibitem{Fink/PRA:1990} M. G. Fink, and W. R. Johnson, Phys. Rev. A 42 (1990) 3801.

\bibitem{Glushkov/PLA:1992} A. V. Glushkov, and L. N. Ivanov, Phys. Lett. A 170 (1992) 33.

\bibitem{Novikov/JPB:2001} S. A. Novikov, and A. N. Hopersky, J. Phys. B 34 (2001) 4857.

\bibitem{Sapirstein/JPB:1996} J. Sapirstein and W. R. Johnson, J. Phys. B 29 (1996) 5213.

\bibitem{Shabaev/PRL:2004} V. M. Shabaev, I. I. Tupitsyn, V. A. Yerokhin, G. Plunien, and G. Soff, Phys. Rev. Lett. 93 (2004) 130405.

\bibitem{Surzhykov/PRA:2010} A. Surzhykov, \textit{et al.}, Phys. Rev. A 81 (2010) 042510.

\bibitem{Volotka/PRA:2011} A. V. Volotka, A. Surzhykov, V. M. Shabaev, and  G. Plunien, Phys. Rev. A 83 (2011) 062508.

\bibitem{Volotka/PRA:2016} A. V. Volotka, V. A. Yerokhin, A. Surzhykov, Th. St\"ohlker, and S. Fritzsche, Phys. Rev. A 93 (2016) 023418.

\bibitem{Salvat/CPC:1995} F. Salvat, J. M. Fernandez-Varea, and W. Williamson Jr., Comput. Phys. Commun. 90 (1995) 151.

\bibitem{Labzowsky/JPB:2005} L. N. Labzowsky, A. V. Shonin, and D. A. Solovyev, J. Phys. B 38 (2005) 265.

\bibitem{Surzhykov/PRA:2009} A. Surzhykov, J. P. Santos, P. Amaro, and P. Indelicato, Phys. Rev. A 80 (2009) 052511.

\end{thebibliography}
\end{document}